\documentclass[aps,prl,onecolumn,superscriptaddress]{revtex4}
\usepackage{graphicx}
\usepackage{movie15}
\usepackage{hyperref}

\begin{document}
\title{Magnetism and its microscopic origin in iron-based high-temperature superconductors}
\author{Pengcheng Dai}
\affiliation{Department of Physics and Astronomy, The University of Tennessee, Knoxville,
Tennessee 37996-1200, USA}
\affiliation{Beijing National Laboratory for Condensed Matter Physics, Institute of
Physics, Chinese Academy of Sciences, P. O. Box 603, Beijing 100190, China}
\author{Jiangping Hu}
\affiliation{Beijing National Laboratory for Condensed Matter Physics, Institute of
Physics, Chinese Academy of Sciences, P. O. Box 603, Beijing 100190, China}  
\affiliation{Department of Physics, Purdue University, West Lafayette, Indiana 47907, USA}
\author{Elbio Dagotto}
\affiliation{Department of Physics and Astronomy, The University of Tennessee, Knoxville,
Tennessee 37996-1200, USA}
\affiliation{Materials Science and Technology Division, Oak Ridge National Laboratory, Oak Ridge, Tennessee 37831, USA}


\begin{abstract}
{\bf High-temperature superconductivity in the iron-based materials emerges from, 
or sometimes coexists with, their metallic or insulating parent compound states. 
This is surprising since these undoped states display
dramatically different antiferromagnetic (AF) spin arrangements and N$\rm \acute{e}$el temperatures.  
Although there is general consensus that magnetic interactions are important 
for superconductivity, much is still unknown concerning the microscopic 
origin of the magnetic states. 
In this review, progress in this area is summarized, focusing on recent 
experimental and theoretical 
results and discussing their microscopic implications. It is concluded 
that the parent compounds are in a state that is more complex than implied by 
a simple Fermi surface nesting scenario, and a dual description including 
both itinerant and localized degrees of freedom is needed to properly 
describe these fascinating materials.}
\end{abstract}

\maketitle

\section{Introduction}

Soon after the discovery of high critical 
temperature (high-$T_c$) superconductivity in copper oxides  \cite{bednorz}, 
neutron scattering studies revealed that 
the parent compounds of these superconductors have an antiferromagnetic 
(AF) ground state with a simple collinear spin structure (Fig. 1a) \cite{vaknin,tranquada88}.  
Because the associated AF spin fluctuations 
may be responsible for electron pairing and superconductivity \cite{scalapino,dagotto,palee}, 
over the past 25 years a tremendous effort has focused on
characterizing the interplay between magnetism and superconductivity in these materials \cite{fujita}.
In the undoped state, the parent compounds of copper oxide superconductors 
are Mott insulators and have exactly one valence fermion with spin 1/2 for each copper atom, 
leading to robust electronic correlations and 
localized magnetic moments \cite{dagotto,palee}.  Superconductivity emerges after introducing charge 
carriers that suppress the static AF order. Although the strong Coulomb repulsion in the parent 
compounds is screened by the doped charge carriers, 
the electronic correlations are certainly important for the physics of the doped cuprates, 
particularly in the underdoped regime \cite{palee}.

Consider now the iron-based superconductors \cite{johnston,stewart,paglione10}. Several 
parent compounds of these materials, such as LaFeAsO, BaFe$_2$As$_2$, NaFeAs, and FeTe, are not
insulators but semimetals \cite{kamihara,rotter,cwchu,mawkuen2}. In these cases, 
electronic band structure calculations have revealed that their Fermi Surfaces (FS) are   
composed of nearly cylindrical 
hole and electron pockets at the $\Gamma(0,0)$ and $M(1,0)/M(0,1)$ points, 
respectively \cite{mazin2011n,hirschfeld}.  
The high density of states resulting from the 
extended momentum space with nearly parallel FS
between the hole and electron pockets leads to an enhancement of the particle-hole susceptibility. This  suggests that 
FS nesting among those pockets
could induce spin-density-wave (SDW) order at the in-plane AF wavevector ${\bf Q}_{AF} = (1,0)$ 
with a collinear spin structure (Fig. 1b) \cite{dong}, much like the FS-nesting induced SDW in
  pure chromium \cite{fawcett}.
Neutron scattering experiments on 
LaFeAsO \cite{cruz}, BaFe$_2$As$_2$ \cite{qhuang}, 
and NaFeAs \cite{slli09} have reported results compatible with the theoretically predicted  
 AF spin structure, albeit with an ordered magnetic moment smaller than expected from
first-principles calculations \cite{mazin08}.
In addition, quasiparticle excitations between the hole and electron FS  
can induce $s^{\pm}$-wave superconductivity \cite{mazin2011n,hirschfeld,kuroki08,chubukov,fwang09}.
One of the consequences of this superconducting state is that the imaginary 
part of the dynamic susceptibility, $\chi^{\prime\prime}(Q,\omega)$ should have a sharp peak,
termed spin resonance in copper oxide superconductors \cite{eschrig}, 
at ${\bf Q}_{AF} = (1,0)$ below $T_c$ \cite{maier,korshunov}.  This prediction is also confirmed by inelastic 
neutron scattering (INS) experiments in iron-based superconductors such as hole-doped Ba$_{1-x}$K$_x$Fe$_2$As$_2$ \cite{christianson,chenglinzhang,castellan}, electron-doped BaFe$_{2-x}T_x$As$_2$ ($T=$Co, Ni) \cite{lumsden,schi09,dsinosov09,jtpark,clester,hfli,hqluo12}, 
and FeTe$_{1-x}$Se$_x$ \cite{hamook,qiu09,lumsden2}.
Finally, angle resolved photoemission spectroscopy (ARPES) experiments find 
that the general characterization of the FS and
the superconducting order parameter are consistent with the band structure 
calculations and with isotropic $s$-wave superconducting gaps \cite{prichard}.
Therefore, at first sight it may appear that antiferromagnetism in the iron-based materials 
originates from FS nesting of itinerant electrons, superconductivity must have 
a $s^{\pm}$-wave symmetry for related reasons, and electron correlations or local moments 
do not play an important role for magnetism and superconductivity \cite{mazin2011n}.

However, although the parent compounds of iron pnictide superconductors 
have metallic ground states consistent with 
band structure calculations, there are reasons to believe 
that electron correlations could be sufficiently strong 
to produce an ``incipient''  Mott physics \cite{si2008,qmsi09}, 
where local moments are as important as itinerant electrons for magnetic, transport, 
and superconducting properties in these materials \cite{cfang,cxu}. 
In fact, the $s^{\pm}$ pairing symmetry is also naturally derived  in 
multi-orbital $t-J$-type models~\cite{seo2008,Fang2011} and recent diagonalization calculations~\cite{nicholson11}
have shown that the AF state, as well as the $A_{1g}$ $s$-wave pairing state, 
evolve smoothly from weak to strong coupling, suggesting
that the physics of the pnictides could also be
rationalized based on short length scale concepts not rooted in weak-coupling nesting.
After all, in the context of the copper oxide superconductors, weak coupling studies of
the one-orbital Hubbard model also led to the correct checkerboard AF state and $d$-wave pairing,
showing that these problems can be attacked from a variety of view points.
In addition,  the newly discovered  $A_y$Fe$_{2-x}$Se$_2$ ($A=$ K, Rb, Cs, Tl)
iron-chalcogenide superconductors~\cite{jgguo,mhfang} do {\it not} 
exhibit hole pockets~\cite{Wang_122Se, Zhang_122Se,Mou_122Se}, 
but have strong AF ordered insulating phases with extremely 
high N\'{e}el transition temperatures~\cite{wbao1,fye}.   
Such a strong magnetism and high superconducting transition temperature ($T_c\approx 33$ K) cannot be explained
by FS nesting since this is based on the enhancement of the particle-hole susceptibility due to an extended
momentum space with nearly parallel Fermi surfaces, i.e. it applies only to particle and hole FS's and not
to purely electronic Fermi pockets.
 
Since iron-based superconductors have  
six electrons occupying the nearly degenerate $3d$ Fe orbitals, the system is intrinsically 
multi-orbital and therefore it is technically difficult to
define and study a simple microscopic Hamiltonian to describe the electronic 
properties of these materials and characterize 
the strength of the electronic correlations. From optical conductivity 
measurements \cite{qazilbash09}, it has been argued that electronic 
correlations in Fe pnictides are weaker than in underdoped copper-oxides 
but are stronger than those of Fermi liquid metals, contrary to the 
conclusion based on local density approximation calculations \cite{mazin2011n}.  Therefore, 
it is important to determine whether magnetism in Fe-based materials arises from weakly correlated
itinerant  electrons~\cite{mazin2011n}, as in the case of the SDW in chromium \cite{fawcett}, or whether it requires some degree of electron
  correlations~\cite{HG}, or if magnetism is dominated by the contributions 
of quasi-localized moments induced by incoherent electronic excitations~\cite{qmsi09} such as in the AF insulating state 
of Cu oxides \cite{palee}.

In this review, recent experimental and theoretical progress in the study of 
iron-based superconductors is summarized, with focus on the undoped parent compounds.  
In section II, the magnetically ordered states in nonsuperconducting
iron pnictides, iron chalcogenides, and iron selenides are discussed.  Section III describes the effect of 
electron and hole doping on static AF order and their associated spin excitations.  In section IV, we provide several examples
where deviations from the simple SDW FS nesting picture are prominent. 
Finally in section V, we present our perspective on  
the importance of electron correlations in these materials.

\section{Magnetic order arrangements and spin waves in the parent compounds}

Although the overall crystal structures and chemical formulas of the copper-oxide superconductors can be quite different, 
their parent compounds are all AF 
Mott insulators characterized by
the Cu spin structure shown in Fig. 1a, where the tetragonal or pseudo-tetragonal unit cells have 
a nearest-neighbor Cu-Cu spacing with $a\approx b\approx 3.8$ \AA.  In the notation 
of reciprocal lattice units (rlu) $(2\pi/a,2\pi/b,2\pi/c)$,
the AF Bragg peaks occur at the in-plane ordering wave vectors ${\bf Q}_{AF}=(\pm 1/2+m,\pm 1/2+n)$, where
$m,n=0,\pm 1, \pm 2, \cdots$ rlu, shown as red circles in Fig.~1e \cite{vaknin,tranquada88}.
Time-of-flight INS experiments \cite{coldea,headings} have mapped out spin waves of 
the insulating La$_2$CuO$_4$ throughout the Brillouin zone and found no evidence for spin-wave broadening at high energies.  
The dispersions of spin waves are well described by a Heisenberg Hamiltonian 
with nearest-neighbor (NN) exchange coupling 
$J_1=111.8\pm 4$ meV and next-nearest-neighbor (NNN) exchange $J_2=-11.4\pm 3$ meV \cite{coldea}.  
Therefore, the dominant magnetic exchange coupling
in La$_2$CuO$_4$ is the NN magnetic interaction and the higher-order interactions amount
to only  $\sim$10\% of the total magnetic energy with a bandwidth of $\sim$320 meV (Fig. 1i).


\begin{figure}[t]
\includegraphics[scale=.7]{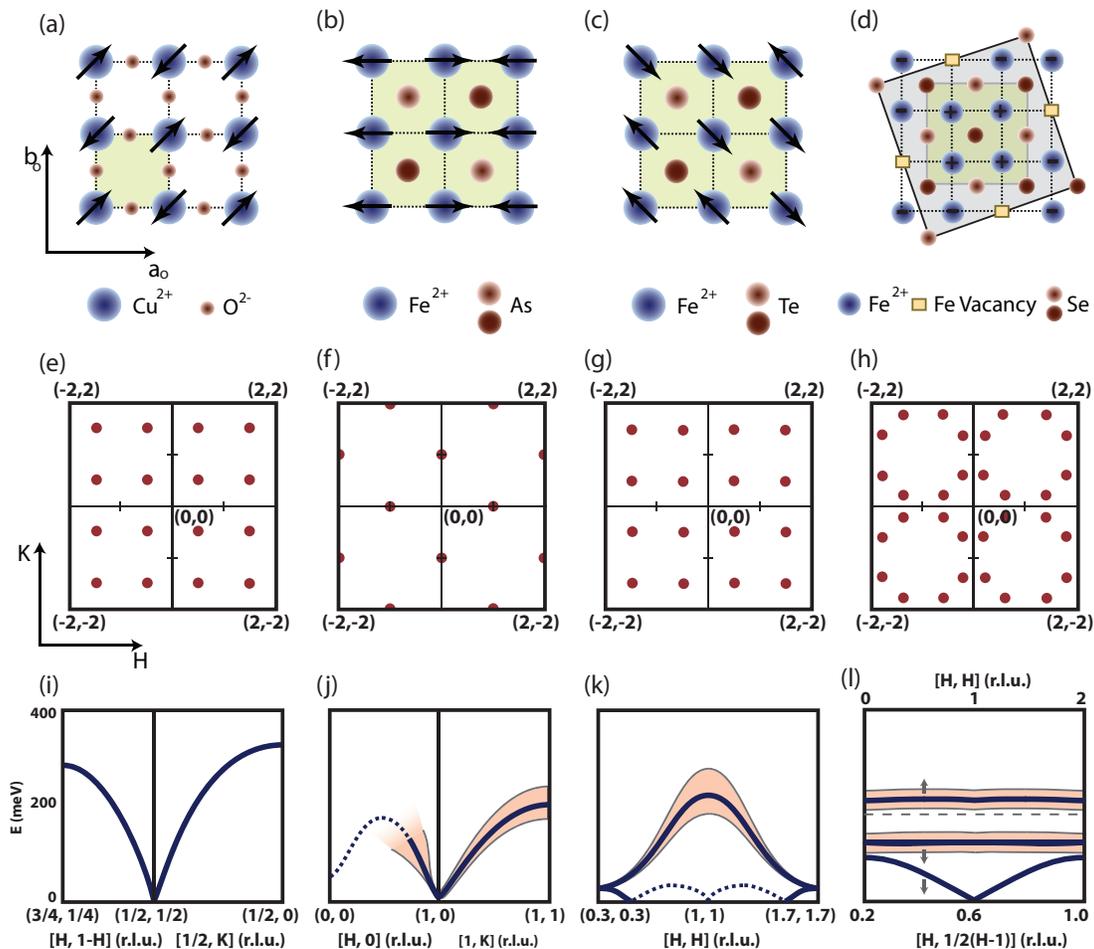}
\caption{Antiferromagnetic structure and spin-wave dispersions for the insulating copper oxide 
La$_2$CuO$_4$ and the parent compounds of iron-based superconductors BaFe$_2$As$_2$, FeTe,
and $A_y$Fe$_{1.6+x}$Se$_2$.  The chemical unit cells are marked light green.  The dark and light brown As/Te/Se atoms indicate their vertical positions above and below the Fe-layer, respectively.
(a) The AF structure of La$_2$CuO$_4$, where the chemical unit cell is marked light green.
(b) The collinear AF structure of nonsuperconducting iron pnictides in the FeAs-layer, where spins are aligned anti-parallel along the orthorhombic $a_o$-axis \cite{cruz,qhuang,slli09}. (c) The bi-collinear AF structure of FeTe \cite{webao08,slli08}. (d) The block AF order of the insulating $A_y$Fe$_{1.6+x}$Se$_2$, where 
the $\sqrt{5}\times\sqrt{5}$ superlattice structure is marked by solid line with lattice parameter $a_s=8.663$ \AA\ and the orthorhombic lattice cell is shaded green \cite{wbao1,fye}.  The iron vacancies are marked as yellow squares. (e) The wave vector dependence of the AF order  
in the $(H,K)$ plane of the reciprocal space for La$_2$CuO$_4$ \cite{coldea}; 
(f) BaFe$_2$As$_2$ \cite{qhuang};
(g) Fe$_{1.05}$Te \cite{webao08,slli08}; and 
(h)  the insulating $A_y$Fe$_{2-x}$Se$_2$ \cite{wbao1,fye}.
(i) Spin-wave dispersions along two high symmetry directions for La$_2$CuO$_4$ \cite{coldea}.  The overall energy
scale of spin waves for copper oxides is about $320$ meV and spin waves are instrumental 
resolution limited. (j) Spin-wave dispersions for BaFe$_2$As$_2$ and they broad considerably for energies 
above $\sim$100 meV \cite{lharriger}.  (k) Spin-wave dispersions for Fe$_{1.05}$Te, and spin waves are very broad for energies above 30 meV \cite{lipscombe}.  
(l) Spin waves for the insulating Rb$_{0.89}$Fe$_{1.58}$Se$_2$ \cite{mywang11}.  In spite of dramatically different dispersions for 
various iron-based materials, their spin wave overall energy scales are similar and about 220 meV, less than that
of the insulating copper oxides. Twinning is considered.}
\end{figure}

Four years after the initial discovery of superconductivity in LaFeAsO$_{1-x}$F$_x$ \cite{kamihara}, 
there are now three major families of 
iron-based superconductors: the iron pnictides \cite{johnston,stewart}, 
iron chalcogenides \cite{mawkuen2,mhfang08}, 
and alkaline iron selenides \cite{jgguo,mhfang}.  The parent compounds of the pnictides, 
such as $A$FeAsO ($A=$ La, Ce, Sm, Pr, etc.), 
$A$Fe$_2$As$_2$ ($A=$ Ba, Sr, Ca), and NaFeAs, all have the same collinear AF structure as shown in Fig.~1b,
with a small ordered moment ($<1\ \mu_B$/Fe) and N$\rm \acute{e}$el temperature $T_N\le 200$ K 
\cite{cruz,qhuang,cwchu}. The AF spin moments  are aligned along the 
weak orthorhombic unit cell $a$-axis direction ($a\approx5.62$, $b\approx 5.57$ \AA). 
In reciprocal space, the AF Bragg peaks occur at in-plane ordering 
wave vectors ${\bf Q}_{AF}=(\pm 1+m, n)$ and at 
${\bf Q}_{AF}\approx (m, \pm 1+n)$ due to twinning (red circles in Fig. 1f), 
consistent with the 
$\Gamma(0,0)\leftrightarrow M(1,0)/M(0,1)$
FS nesting picture \cite{mazin2011n}.   
However, although the calculated FS of the chalcogenides Fe$_{1+y}$Te$_{1-x}$Se$_x$ 
is similar to that of iron pnictides \cite{subedi}, surprisingly its parent compound
Fe$_{1+y}$Te actually has a bi-collinear spin structure 
(Fig. 1c) \cite{webao08,slli08}.  Here, the AF Bragg peaks occur 
at ${\bf Q}_{AF}=(\pm 1/2+m,\pm 1/2+n)$ (Fig. 1g) in 
the pseudo-tetragonal notation ($a\approx b\approx 5.41$ \AA),  suggesting that
FS nesting cannot induce such AF order.  Finally, the parent compounds of the
alkaline iron selenide $A$Fe$_{1.6+x}$Se$_2$ superconductors 
are insulators \cite{jgguo,mhfang}
and form a $\sqrt{5}\times\sqrt{5}$ block AF structure as shown in Fig. 1d 
with a large ordered moment ($\sim$$3\ \mu_B$/Fe) along the $c$-axis and
$T_N\approx 500$ K \cite{wbao1,fye}.  
 In reciprocal space, defined using the pseudo-tetragonal unit cell 
of iron pnictides ($a\approx b\approx 5.41$ \AA),
the block AF Bragg peaks appear at ${\bf Q}_{AF}=(\pm 0.2+m,\pm 0.6+n)$ and 
$(\pm 0.6+m,\pm 0.2+n)$ combining left and right chiralities (red circles in Fig. 1h).

Since the parent compounds of iron-based superconductors can have 
different AF spin structures and either
metallic or insulating ground states \cite{johnston,stewart,jgguo,mhfang}, 
the microscopic origin of the
AF order cannot be induced by a simple FS nesting. 
If magnetism is relevant for high-$T_c$ superconductivity, 
then it would be important to determine magnetic exchange couplings for different 
 classes of Fe-based superconductors and compare the results with those 
of the copper-oxides \cite{coldea}. For pnictides, INS
experiments have mapped out spin waves 
on single crystals of CaFe$_2$As$_2$ \cite{diallo09,jzhao}, SrFe$_2$As$_2$ \cite{raewings}, 
and BaFe$_2$As$_2$ \cite{lharriger} throughout the Brillouin zone.  Although there are still 
debates concerning whether spin waves in
these materials can be described by a pure itinerant picture \cite{diallo09,raewings} or require 
local moments \cite{jzhao,lharriger}, the overall spin-wave energy scales  
are around 220 meV.  
Therefore, magnetic exchange couplings in iron pnictides are clearly smaller than those of copper 
oxides (Figs. 1i and 1j). Although spin waves are broadened at high energies, the
spin-wave dispersion curves (Fig. 1j)  can still be described by a Heisenberg Hamiltonian with 
strong anisotropic NN exchange couplings ($J_{1a}\gg J_{1b}$) and fairly large NNN exchange 
coupling ($J_2$) \cite{jzhao,lharriger}. This large in-plane magnetic exchange coupling 
anisotropy has been interpreted as due to possible electronic nematic 
phase and/or orbital ordering \cite{jzhao,lharriger}. Table I compares the effective magnetic exchange 
couplings of the Fe-based systems studied thus far against those of the insulating copper-oxide La$_2$CuO$_4$.

For the chalcogenides Fe$_{1+y}$Te, the commensurate bi-collinear AF spin structure in Fig. 1c 
becomes incommensurate for concentration $y>0.12$ \cite{rodriguez}.  
The overall spin-waves energy scale (Fig. 1k) is similar to those of the iron pnictides.
Although the large static ordered moment of $\sim$$2\ \mu_B$/Fe in Fe$_{1+y}$Te \cite{webao08,slli08} 
suggests that local moments may be important, spin waves are rather broad in energy 
and difficult to fit using a Heisenberg Hamiltonian with only NN and NNN exchange couplings \cite{lipscombe}.
By including third-neighbor (NNNN) exchange couplings, a Heisenberg Hamiltonian can fit the
spin-wave dispersion with an anisotropic ferromagnetic NN exchange 
couplings and strong AF NNN exchange coupling (Table I). In a separate INS 
 experiment on  Fe$_{1.1}$Te, the total integrated Fe magnetic moment was
 found to increase with increasing temperature from 10 K to 80 K \cite{zaliznyak}.  
These results suggest that in the temperature range relevant for superconductivity, 
there is a remarkable redistribution of the magnetism arising from both 
itinerant and localized electrons.

\begin{figure}[t]
\includegraphics[scale=.55]{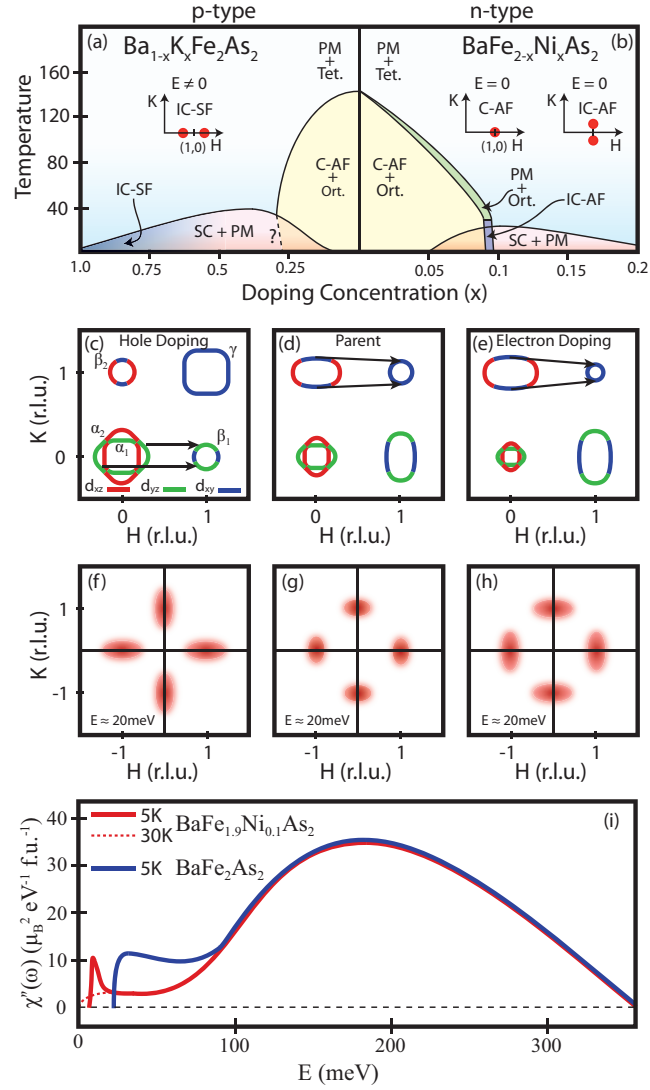}
\caption{The electronic phase diagrams and the evolution of
FS's, static AF order, and spin excitations upon electron or hole doping to BaFe$_2$As$_2$.
(a) The AF and superconducting phase diagram for hole-doped Ba$_{1-x}$K$_x$Fe$_2$As$_2$.  In the underdoped
regime, there is a region of coexisting AF order and superconductivity \cite{ewiesenmayer11}.  
Incommensurate spin excitations appear for $x\ge 0.4$ \cite{castellan} and persist till $x=1$ 
at KFe$_2$As$_2$ \cite{chlee11}. (b) Phase diagram for electron-doped 
BaFe$_{2-x}$Ni$_x$As$_2$ \cite{hluo12}.  The long commensurate AF order changes into short-range incommensurate 
AF order near $x=0.092$. The inset shows the transverse incommensurate AF order. Superconductivity 
in the electron-doped materials only extends to $x\approx 0.25$. 
(c) Schematics of FS's correspond to 35\%
hole-doped BaFe$_2$As$_2$ \cite{chenglinzhang} with possible nesting vectors marked with arrows.  
The $d_{xz}$, $d_{yz}$, and $d_{xy}$ orbitals for different Fermi 
surfaces are colored as red, green, and blue, respectively. (d) FS's of BaFe$_2$As$_2$ with
orbital characters \cite{graser10}. (e) Fermi surfaces of 8\% electron-doped 
BaFe$_2$As$_2$ \cite{chenglinzhang}.  For all three cases, 
FS's are plotted with zero wave vector transfers along the $c$-axis.
(f) Longitudinally elongated spin excitations at $E=20$ meV seen in the
optimally hole-doped Ba$_{0.67}$K$_{0.33}$Fe$_2$As$_2$ \cite{chenglinzhang}. 
(g) Transversely elongated spin waves at $E=20$ meV for BaFe$_2$As$_2$  \cite{lharriger}. 
(h) Transversely elongated spin excitations at $E=20$ meV for BaFe$_{1.9}$Ni$_{0.1}$As$_2$ \cite{jtpark,msliu12}. (i) Energy dependence of $\chi^{\prime\prime}(\omega)$ for BaFe$_2$As$_2$ (blue solid line)
and BaFe$_{1.9}$Ni$_{0.1}$As$_2$ below (red dashed line) and above (red solid circles) $T_c$ in absolute units of
$\mu_B^2$eV$^{-1}$f.u.$^{-1}$. The sharp peak near $E\approx 8$ meV below $T_c$ is the neutron spin resonance coupled directly to superconductivity \cite{lumsden,schi09,dsinosov09,jtpark,clester,hfli,hqluo12}.
}
\end{figure}

In the case of insulating $A$Fe$_{1.6+x}$Se$_2$, spin waves have an acoustic mode 
and two optical modes separated by spin gaps (Fig. 1l) \cite{mywang11}. 
In contrast to iron pnictide $A$Fe$_2$As$_2$ \cite{diallo09,jzhao,raewings,lharriger} and
iron chalcogenide Fe$_{1+y}$Te \cite{lipscombe,zaliznyak}, spin waves in insulating $A$Fe$_{1.6+x}$Se$_2$
can be well-described by a Heisenberg Hamiltonian with NN, NNN, and NNNN exchange couplings \cite{mywang11}. Comparing effective exchange couplings for different iron-based materials (Table I), 
it is clear that the NN exchange couplings are quite different, but the NNN exchange couplings are AF and rather similar. In addition, spin waves for iron-based materials are much broader at high energies.
This is different from the insulating copper oxides, where the NN exchange coupling dominates the magnetic interactions and spin waves are instrumental resolution limited throughout the Brillouin zone \cite{coldea,headings}.  These results suggest that itinerant electrons play a role in spin waves of metallic iron-based materials.

\begin{table}[h]
\begin{center}
     \begin{tabular}{ |l | l | l | l | l | l | l |}
     \hline
           Parent compounds  & $T_N$ (K)    & $SJ_{1a}$ (meV)       & $SJ_{1b}$ (meV)       & $SJ_{2a}$ (meV)         & $SJ_{2b}$ (meV) &  $SJ_{3}$ (meV) \\ \hline
La$_2$CuO$_4$, Ref. \cite{coldea} & $317\pm 3$ & $111.8\pm4$ & $111.8\pm4$ & $-11.4\pm 3$ & $-11.4\pm 3$  & 0 \\ \hline
CaFe$_2$As$_2$, Ref. \cite{jzhao} & $\sim$170 & $49.9\pm 9.9$ & $-5.7\pm4.5$ & $18.9\pm 3.4$ & $18.9\pm 3.4$ & 0 \\
\hline
BaFe$_2$As$_2$, Ref. \cite{lharriger} & $\sim$138 & $59.2\pm2.0$ & $-9.2\pm 1.2$ & $13.6\pm 1$ & $13.6\pm 1$ & 0 \\
\hline
Fe$_{1.05}$Te, Ref. \cite{lipscombe} & $\sim$68 & $-17.5\pm 5.7$ & $-51.0\pm 3.4$ & $21.7\pm 3.5$ & $21.7\pm 3.5$ & $6.8\pm 2.8$ \\
\hline
Rb$_{0.89}$Fe$_{1.58}$Se$_2$, Ref. \cite{mywang11} & $\sim$475 & $-36\pm 2$ & $15\pm 8$ & $12\pm 2$ & $16\pm 5$ & $9\pm 5$ \\
\hline
     \end{tabular}
     \caption{\label{tab:5/tc}
Comparison of effective magnetic exchange couplings for parent compounds of copper-based 
and iron-based superconductors obtained by fitting spin waves with a Heisenberg Hamiltonian with 
NN ($J_{1a}, J_{1b}$), NNN ($J_{2a},J_{2b}$), and NNNN ($J_3$).  
The N$\rm \acute{e}$el temperatures for different materials are also listed.     
     }
     
 \end{center}
\end{table}

\section{The effects of hole and electron doping on the magnetic correlations and excitations}

As discussed before  \cite{johnston,stewart,paglione10}, superconductivity 
in Fe-based materials can be induced via electron/hole doping, pressure, and isoelectronic substitution.
Figures 2a and 2b show the electronic phase diagrams of
hole and electron  doping on BaFe$_2$As$_2$, respectively.  
In the undoped state, BaFe$_2$As$_2$ exhibits simultaneous structural and magnetic phase transitions below $\sim$138 K, changing
from the high-temperature paramagnetic tetragonal phase to the low-temperature orthorhombic phase with the collinear AF structure (Fig. 1b) \cite{qhuang}.
Upon electron-doping BaFe$_2$As$_2$ by partially replacing Fe by Co or Ni to form BaFe$_{2-x}T_x$As$_2$, the static AF order is suppressed and 
superconductivity emerges \cite{johnston,stewart,paglione10}.  From systematic transport and magnetic measurements of 
single crystals \cite{nni08,jhchu09}, the phase diagram for BaFe$_{2-x}$Co$_x$As$_2$ was established, where the single 
structural/magnetic phase transition in BaFe$_2$As$_2$ splits with increasing Co-doping.  Neutron diffraction experiments 
on BaFe$_{2-x}$Co$_x$As$_2$ \cite{lester09} confirm
that the commensurate AF order appears below the structural transition 
temperature and superconductivity coexists with AF order for $0.06\leq x\leq 0.102$.
Neutron scattering measurements on BaFe$_{2-x}$Co$_x$As$_2$ 
with coexisting AF order and superconductivity reveal that the intensity of AF Bragg peaks
actually decreases below $T_c$ without changing the spin-spin correlation lengths \cite{pratt09,adchristianson09}.  While these results indicate 
that the static AF order competes with superconductivity, it remains unclear whether the long-range AF order truly coexists microscopically 
with superconducting regions \cite{mywang,mywang11b}.  Recently, 
for electron-doped samples near optimal superconductivity it has been shown that
 the commensurate
static AF order changes into transversely incommensurate short-range AF order that coexists and competes with superconductivity 
(see inset in Fig. 2b) \cite{dkpratt11,hluo12}.  Taking the temperature dependence of the orthorhombic 
lattice distortion of BaFe$_{2-x}$Co$_x$As$_2$ into account \cite{snandi10}, the AF order, structure, and 
superconductivity phase diagrams for BaFe$_{2-x}T_x$As$_2$ are shown in Fig.~2b.

Although the superconducting transition temperature for hole-doped 
Ba$_{1-x}$K$_x$Fe$_2$As$_2$ can reach up to $T_c=38$ K \cite{rotter} as compared to the
$T_c\approx 25$ K for electron-doped BaFe$_{2-x}T_x$As$_2$ \cite{johnston,stewart}, those materials
are much less studied because of the difficulty in growing 
high-quality single crystals. The initial transport and neutron scattering  
experiments on powder samples indicated a gradual suppression of the concurrent structural and
magnetic phase transitions with increasing K-doping.  For the underdoped regime 
$0.2\leq x\leq 0.4$, commensurate AF order appears to microscopically coexist with
superconductivity \cite{hchen09}.  Subsequent neutron scattering
and muon spin rotation ($\mu$SR) measurements on 
single crystals grown in Sn-flux suggested mesoscopic separation of the AF and superconducting phases \cite{jtpark09}.  However, recent  
neutron \cite{savci}, X-ray scattering, and $\mu$SR work \cite{ewiesenmayer11} on high-quality powder samples confirm the microscopic coexistence of the commensurate AF order with superconductivity in the underdoped region between $0.2\leq x\leq 0.3$ and the suppression of the
orthorhombic phase below $T_c$ (Fig. 2a). 
Since currently there is no neutron diffraction work on high-quality single crystals of Ba$_{1-x}$K$_x$Fe$_2$As$_2$ grown using FeAs-flux, it is unclear if there are also short-range incommensurate AF order in Ba$_{1-x}$K$_x$Fe$_2$As$_2$ near optimal superconductivity.

The appearance of static incommensurate AF order along the transverse direction of 
the collinear AF ordering wave vector ${\bf Q}_{AF}=(\pm 1,0)$
in BaFe$_{2-x}T_x$As$_2$ suggests that such order arises from the electron doping effect of 
FS nesting \cite{dkpratt11,hluo12}. Based on 
a five-orbital tight-binding model, fitted to
the density functional theory (DFT) band structure for BaFe$_2$As$_2$ \cite{graser10}, 
there should be five FS pockets with different orbital contributions in the two-dimensional reciprocal space at ${\bf Q}_z=0$ (Fig. 2d). 
The intraorbital, but interband, scattering process between $\Gamma(0,0)\leftrightarrow M(1,0)$ shown in Fig. 2d favors 
the transversely lengthened vertices \cite{jhzhang10}. This momentum anisotropy is compatible with the experimentally observed 
elliptically shaped low-energy spin excitations in superconducting BaFe$_{2-x}T_x$As$_2$ \cite{jtpark,clester,hfli,hqluo12} and spin waves
in BaFe$_2$As$_2$ (Fig. 2g) \cite{lharriger}.  Upon electron-doping to enlarge the electron pockets near 
$M(1,0)/(0,1)$ and shrink the hole pockets near $\Gamma(0,0)$,  the mismatch  
between the electron and hole Fermi pockets
becomes larger (Figs. 2e), resulting in a more transversely elongated ellipse in the low-energy
 magnetic response (Fig. 2h).  Indeed, this is qualitatively 
 consistent with the doping evolution of the low-energy spin excitations \cite{jtpark,hqluo12,msliu12}.

For hole-doped Ba$_{1-x}$K$_x$Fe$_2$As$_2$, one should expect enlarged hole Fermi pockets near $\Gamma(0,0)$ and reduced electron pockets near 
$M(1,0)/(0,1)$, as shown in Fig. 2c.  
Based on first principles calculations, spin excitations for optimally hole-doped 
Ba$_{1-x}$K$_x$Fe$_2$As$_2$ at $x=0.4$ should have longitudinally elongated ellipses \cite{jtpark}, 
and gradually evolve into incommensurate magnetic scattering (elastic and/or inelastic) with increasing $x$ 
due to poor nesting between the hole and electron Fermi pockets \cite{castellan}. 
INS experiments on single crystal 
Ba$_{0.67}$K$_{0.33}$Fe$_2$As$_2$ \cite{chenglinzhang} indeed confirm that the low-energy spin excitations 
are longitudinally elongated ellipses that are rotated 90$^\circ$ from
that of the electron-doped BaFe$_{2-x}T_x$As$_2$ (Fig. 2f) \cite{jtpark,clester,hfli,hqluo12}. Furthermore, INS
measurements on powder samples of Ba$_{1-x}$K$_x$Fe$_2$As$_2$ reveal that spin excitations change from commensurate 
to incommensurate for $x\ge 0.4$, although their exact line shape and incommensurability in reciprocal space are unknown \cite{castellan}.
Finally, INS experiments on hole-overdoped KFe$_2$As$_2$ found incommensurate spin fluctuations along 
the longitudinal direction (inset in Fig.~2a), again consistent with the FS nesting picture \cite{chlee11}. 
Figure 2a shows the electronic phase diagram of hole-doped Ba$_{1-x}$K$_x$Fe$_2$As$_2$ based on the present understanding of these materials.

Although FS nesting is compatible with a number of experimental observations 
for the evolution of spin excitations in electron/hole-doped iron-based 
superconductors, there are several
aspects of the problem where such a scenario cannot be reconciled with experiments.  
In a recent INS experiment on optimally electron-doped  
BaFe$_{1.9}$Ni$_{0.1}$As$_2$, magnetic excitations throughout the Brillouin zone have been measured in absolute units and compared with spin waves for
AF BaFe$_2$As$_2$  \cite{msliu12}. In the fully localized (insulating) case, the formal
Fe$^{2+}$ oxidation state in BaFe$_2$As$_2$ would give a $3d^6$ electronic configuration and 
Hund's rules would yield $S=2$.  The   
total fluctuating moments should be $\left\langle m^2\right\rangle=(g\mu_B)^2
S(S+1)=24\ \mu_B^2$ per Fe assuming $g=2$ \cite{clester,msliu12}.  For spin waves in the insulating Rb$_{0.89}$Fe$_{1.58}$Se$_2$, the total moment sum rule 
appears to be satisfied \cite{mywang11}.  The fluctuating moments for BaFe$_2$As$_2$ and BaFe$_{1.9}$Ni$_{0.1}$As$_{2}$ are  $\left\langle
m^2\right\rangle=3.17\pm 0.16$ and $3.2\pm 0.16\ \mu_B^2$ per Fe(Ni), respectively \cite{msliu12}.  While these values 
are considerably smaller than those of the fully localized case, 
they are much larger than expected from the fully itinerant SDW using the random phase approximation \cite{hpark12}. A calculation combining DFT 
and dynamical mean field theory (DMFT)
suggests that both the band structure and the local 
moment aspects (e.g. Hunds coupling) of the iron electrons are needed for a good description of the magnetic responses \cite{msliu12}.  Figure 2i 
shows the energy dependence of $\chi^{\prime\prime}(\omega)$ for BaFe$_2$As$_2$ and BaFe$_{1.9}$Ni$_{0.1}$As$_{2}$, and it is clear that 
the impact of electron doping and superconductivity are limited to spin excitation energies below 100 meV.  These results suggest that high-energy spin excitations are likely to arise from the 
local moments instead of FS nesting effects.

\section{Deviations from the simple SDW Fermi surface nesting picture}

After the early research efforts on Fe-based 
superconductors~\cite{johnston,stewart,paglione10}, recent experimental 
and theoretical investigations are providing a more refined perspective of 
these materials. Below, 
several selected examples will be discussed, supplementing
those presented in the neutrons sections.

{\it Strength of electronic correlations}. The strength of electronic correlations 
are often characterized via the ratio 
between the on-site Hubbard repulsion coupling $U$ and the bandwidth $W$ of the 
hole or electron carriers.
Early on, it was assumed that pnictides were in the weak-interaction 
limit $U/W\ll 1$. However, recent investigations revealed that the electronic 
correlations induce large enhancements between the
effective and bare electronic masses, signaling that 
correlation effects cannot be neglected. For instance, 
Haas-van Alphen experiments for 
KFe$_2$As$_2$ unveiled discrepancies between the band-theory 
calculated and observed FS's, including a large electronic mass 
enhancement 3-7 caused by band narrowing \cite{tterashima10}. Similar ratios 
for the overdoped Tl$_2$Ba$_2$CuO$_{6+\delta}$ copper-oxides have been 
reported~\cite{pmcrourke}, suggesting that the undoped parent compounds of the pnictides 
resemble the overdoped copper oxides.

Additional insight is provided by optical conductivity experiments, 
since the ratio $R$ between the experimentally measured kinetic energy 
and that of band-theory calculations can be measured and 
contrasted against other compounds \cite{qazilbash09}. $R\approx 1$ signals 
a good metal such as Ag. LaFePO presents a ratio $R\approx 0.5$ which is borderline 
between weak and moderate coupling. However, pnictides such as BaFe$_{2-x}T_x$As$_2$  
are characterized by an even stronger correlation that induces a ratio $R\approx 0.3$ 
which is similar to results for overdoped La$_{2-x}$Sr$_x$CuO$_4$, widely considered to be a ``correlated metal''.  
Other studies have arrived to similar conclusions with regards 
to the correlation strength \cite{nakamura,dsinosov11}. In agreement 
with experiments, DFT+DMFT predicts a mass enhancement 
$m^\ast/m_{band}\sim$2-3 for BaFe$_{2-x}T_x$As$_2$ and $\sim$7 
for FeTe \cite{zpyin10}. Moreover, ARPES studies 
of NaFeAs revealed band reconstructions in the magnetic state involving 
bands well below the FS~\cite{che10}, contrary to a weak coupling picture.

\begin{figure}[t]
\includegraphics[scale=.48]{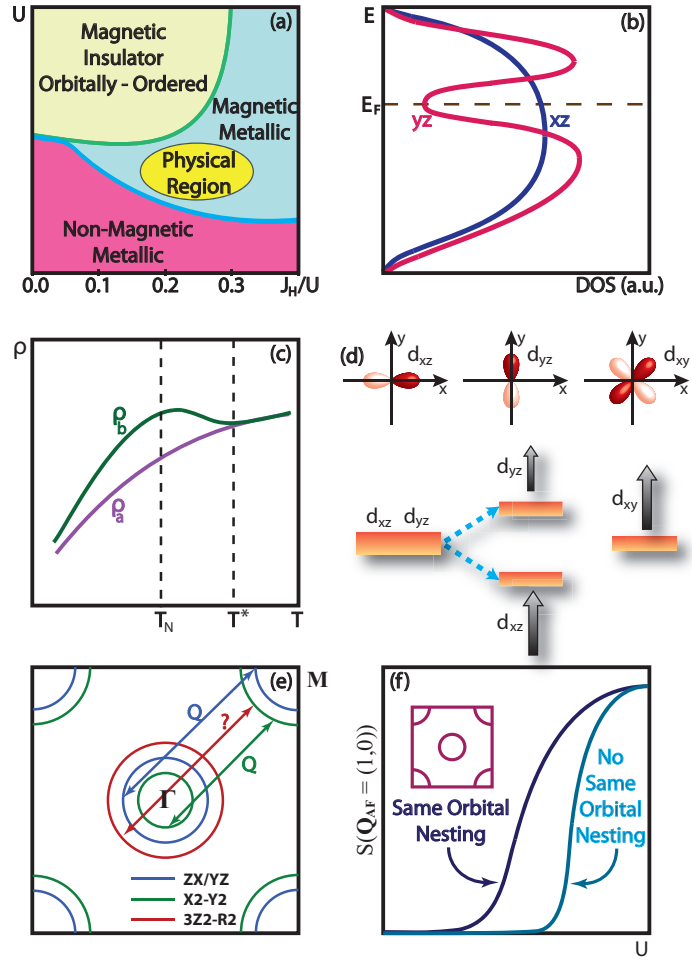}
\caption{Summary of the phase diagram of multiorbital Hubbard models 
and the electronic state of Fe near the FS.
(a) Sketch of the phase diagram of a typical multiorbital Hubbard model 
in the undoped limit, varying the on-site same-orbital repulsion $U$ and 
the ratio between the Hund coupling $J_H$ and $U$.  Highlighted is a region 
dubbed ``physical region'' where the properties of the model are in good 
agreement with experiments. Note the location of this region in the intermediate 
magnetic-metallic phase, with magnetic order at ${\bf Q}_{AF}=(1,0)$, at similar 
distance from the paramagnetic state and from the insulator orbitally-ordered state \cite{qluo10}.
(b) Sketch of the DOS illustrating the phenomenon of FS orbital order, which 
is a weight redistribution at the FS of the states associated with the $xz$ 
and $yz$ $d$-orbitals. Even though the integral over energy gives similar values 
for both orbitals, at the FS there are drastic differences that influence 
on several properties such as transport \cite{daghofer10,daghofer12}.
(c) Sketch of the anisotropy found in transport experiments for detwinned 
Ba(Fe$_{1-x}$Co$_x$)$_2$As$_2$. Note that this anisotropy is present 
at temperatures substantially larger than $T_N$ \cite{irfisher11}.
(d) Orbitals of relevance for the discussion of the Fe-based superconductors 
and their splitting at the FS.
(e) Sketch of the ARPES results of Ref. \cite{shimojima11}, illustrating 
the absence of a nesting partner for one of the hole pockets. The material 
still displays a nearly uniform superconducting gap at this and all the other 
hole and electron pockets.
(f) Sketch of the magnetic moment at wave vector ${\bf Q}_{AF}=(1,0)$ for 
two models. On the left is the result for a traditional model of pnictides, 
with the $xz$, $yz$, and $xy$ $d$-orbitals active at the FS. 
This model displays magnetic order in a broad range of couplings from very 
weak to strong. On the right, results of a model with the same FS but totally 
different orbital composition. While at small 
$U$ there is no order, at larger couplings this model 
converges to the same ${\bf Q}_{AF}=(1,0)$ order \cite{nicholson}.
}
\end{figure}

Hubbard model investigations provide additional insight 
on this subject. When compared with similar efforts 
for the cuprates, the study of Hubbard models for the pnictides 
is far more challenging because several Fe orbitals are needed. For this reason, many efforts are restricted to 
mean-field Hartree-Fock approximations. For the undoped three-orbital 
Hubbard model, employing the $d_{xz}$, $d_{yz}$, and $d_{xy}$ orbitals 
of relevance at the FS, a sketch of a typical mean-field 
phase diagram varying $U$ and the Hund coupling $J_H$ \cite{qluo10} is in Fig.~3a. 
Three regimes are identified: a small $U$ phase where the state is paramagnetic, 
followed with increasing $U$ by an intermediate regime 
simultaneously metallic and magnetic~\cite{qluo10}, and finally a large-$U$ phase 
where a gap in the 
density-of-states (DOS) is induced leading to an insulator (with concomitant orbital order). Comparing 
the theoretical predictions for the magnetic moment in the ${\bf Q}_{AF}=(1,0)$ 
wave-vector channel against neutrons, and the one-particle spectral function 
$A({\bf k},\omega)$ against ARPES, the intermediate-coupling region dubbed 
``physical region'' in yellow in Fig.~3a represents qualitatively 
the undoped BaFe$_{2-x}T_x$As$_2$ compounds \cite{qluo10}. In this regime, 
$U/W\sim$0.3-0.4, and similar results were reported for the two- and five-orbital models~\cite{mDaghofer}. 
Note that Hartree-Fock usually produces critical couplings 
smaller than they truly are because of the neglect of quantum fluctuations.  
In fact, recent investigations beyond Hartree-Fock ~\cite{kkubo11} 
suggest that the relevant $U$ may be larger than those found in Hartree-Fock \cite{qluo10} by 
approximately a factor two. The study of
effective low-energy Hamiltonians starting from first-principles calculations
also led to the conclusion that $U/W$ is between 0.5 and 1.0 for the pnictides depending on the particular
compound~\cite{arita-imada}. Thus, 
the regime of relevance is neither very weak coupling nor strong 
coupling but the more subtle, and far less explored, intermediate  region. 
Previous efforts converged to similar conclusions~\cite{Johannes09}. 
This is also compatible with the notion
that the parent compound is close to a Mott insulator~\cite{si2008,qmsi09}.  
In the ``physical region'' the ratio $J_H/U$ is approximately 1/4 ~\cite{qluo10}, 
as in other estimations \cite{zpyin10}, highlighting the 
importance of $J_H$ in these materials that are sometimes referred to as Hund metals~\cite{hund-metal}. 
Finally, it is very important to remark that
the above described analysis of $U/W$ holds for pnictides but the recent discovery of the alkaline
iron selenides \cite{jgguo,mhfang} has opened a new chapter in this field and it is
conceivable that for these materials $U/W$ will be larger than in pnictides explaining, for
example, the large values of the iron moments.

{\it Role of the orbital degree of freedom}. 
The ``physical region'' in Fig.~3a is not only close to the paramagnetic regime, 
but also similarly close to the insulator, which in the mean-field approximation 
is also orbitally ordered \cite{mDaghofer}. The potential relevance of the orbital 
degree of freedom in pnictides has been discussed \cite{lv10,wgyin10}. 
The orbital can be of relevance not only in its long-range-ordered 
form, but also via its coupling to the spin and its influence near the FS. 
In fact, polarized ARPES experiments on BaFe$_2$As$_2$ \cite{shimojima} reported that 
at the FS there was an asymmetry between the populations of the $d_{xz}$ 
and $d_{yz}$ orbitals. Theoretical studies showed that this effect indeed 
occurs in the ${\bf Q}_{AF}=(1,0)$ magnetic state, and it is linked to an 
orbital-dependent reduction in the DOS at 
the FS \cite{daghofer10}, sketched in Fig. 3b, phenomenon
dubbed ``Fermi surface orbital order''.

This effect, while not sufficiently strong to induce long-range order 
as in manganites, can still severely influence the 
properties of the material. Consider for example the transport anisotropy 
observed in detwinned BaFe$_{2-x}T_x$As$_2$ single-crystals \cite{irfisher11,matanatar}, 
sketched in Fig. 3c. At low temperatures the difference between the 
$a$-axis (spins antiparallel, Fig. 1b) and $b$-axis (spin parallel, Fig. 1b)
directions can be rationalized based on the magnetic state, since 
the different spin arrangements along the $a$ and $b$ break rotational 
invariance~\cite{xtzhang}. However, both in the undoped case and particularly 
in the lightly-doped regime, the asymmetry persists well above the N$\rm \acute{e}$el 
temperature, $T_N$, into a new temperature scale $T^\ast$ that may be associated 
with the onset of nematic order \cite{cfang,cxu}, similarly as 
in some ruthenates and copper oxides \cite{efradkin}.  ARPES experiments 
on the same materials \cite{myi} reported a $d_{xz}$ and $d_{yz}$ band 
splitting (Fig.~3d) that occurs above $T_N$ in the same region where transport 
anisotropies were found. Although the splitting is too small to be a canonical 
long-range orbital order, it reveals the importance of fluctuations above the 
critical temperatures. Optical spectra studies also unveiled anisotropies in 
the spectra persisting up to 2 eV, incompatible with SDW 
scenarios \cite{Nakajima}. Note that the discussion on this subject is still 
fluid. While neutron diffraction investigations showed that $T_N$ actually 
substantially increases as the pressure needed to detwin the crystals increases, 
potentially explaining the observed resistivity anisotropies \cite{dhital}, 
magnetic torque measurements without external pressure revealed clear evidence for 
 electronic nematicity \cite{kasahara}.
Recent calculations addressing transport indeed find an important role of the orbital 
states above $T_N$ \cite{rmfernandes}. The
orbital degree of freedom, closely entangled to the spin and the lattice, may lead 
to a more complex ``normal'' state than anticipated from weak coupling
particularly because of the FS orbital order~\cite{daghofer10}. 
In fact, neutron scattering shows that although the low-energy magnetic 
dispersion changes substantially when crossing critical temperatures, the higher 
energy features remain the same over a large doping and temperature range \cite{msliu12}, 
suggesting that spin, orbital, and lattice are closely entangled. Establishing who 
is the ``driver'' and who is the ``passenger'' may define an important area of 
focus of future research.

{\it Local moments at room temperature}. Another deviation from a simple weak coupling picture 
is the observation of local magnetic moments at room temperature. 
Within the SDW scenario, magnetic moments are formed upon 
cooling simultaneously with the development of long-range magnetic order.
But recent Fe X-ray emission 
spectroscopy experiments unveiled the existence of local moments in the 
room-temperature paramagnetic state~\cite{hgretarsson}. In fact, with the only 
exception of FeCrAs, for all the pnictides 
and chalcogenides investigated a sizable room temperature magnetic moment was found. 
This includes LiFeAs, that actually does not order magnetically at any 
temperature~\cite{cwchu}, and  $A$Fe$_{1.6+x}$Se$_2$ with 
a regular arrangement of Fe vacancies (Fig. 1d). These observed local moments 
are similar in magnitude to those reported in the low-temperature neutron 
scattering experiments reviewed in previous sections. Similar conclusions 
to those of~\cite{hgretarsson} were reached in a study of $3s$ core level 
emission for CeFeAsO$_{0.89}$F$_{0.11}$~\cite{bondino} and also in LDA+DMFT 
investigations~\cite{hansmann}.

{\it Polarized ARPES results and orbital composition}. 
While research using ARPES techniques applied to pnictides has 
already been reviewed~\cite{prichard},
 some intriguing recent results addressing the influence of nesting 
are included in our discussion. Using bulk-sensitive laser ARPES on 
BaFe$_2$(As$_{0.65}$P$_{0.35}$)$_2$ and Ba$_{0.6}$K$_{0.4}$Fe$_2$As$_2$, 
an orbital-independent superconducting gap magnitude was found for the 
hole-pockets FS's \cite{shimojima11}. These results are 
incompatible with nesting where the FS nested portions 
must have a robust component of the same orbital to be effective. 
Actually, the red hole pocket shown in the sketch in Fig. 3e, that experimentally 
displays a robust and nearly wave-vector-independent superconducting gap similar to those 
found in the other hole pockets, does not have a matching electron pocket with the 
same orbital composition and, thus, it cannot develop its superconductivity via a 
nesting pairing mechanism \cite{mazin2011n}. Perhaps inter-orbital 
pairing~\cite{amoreo} or orbital fluctuations could be relevant to explain 
this paradox. Recent theoretical work~\cite{nicholson} addressed the importance 
of orbital composition via two models: one with nested electron- and 
hole-pocket Fermi surfaces with the standard orbital composition of pnictide models, 
and another with 
the same FS shape but with electron and hole pockets having totally different 
orbital compositions. As sketched in Fig. 3f, the former develops magnetic order 
at smaller values of $U$ than the latter. However, 
 with sufficiently large $U$ both have magnetic ground states with the same 
wavevector ${\bf Q}_{AF}=(1,0)$ (Fig. 3f). At large $U$ it is clear that 
the ${\bf Q}_{AF}=(1,0)$ 
order can be understood within a local picture, 
based on the similar magnitude of the super-exchange 
interactions between NN and NNN spins using a simple 
Heisenberg model.

{\it Additional experimental results}. De Haas-van Alphen 
studies \cite{bjarnold} in non-superconducting BaFe$_2$P$_2$, the end member of the 
superconducting series BaFe$_2$(As$_{1-x}$P$_x$)$_2$, indicate that the differences 
in the pairing susceptibility varying $x$ are caused by increases in 
$U$ and $J_H$ rather than improved geometric 
nesting. Moreover, ARPES studies of LiFeAs, without
long-range magnetic order at low temperatures, report a strong renormalization 
of the band structure by a factor $\sim$3 and the absence of 
nesting \cite{Borisenko}. Yet, at $T_c=18$ K \cite{cwchu} LiFeAs 
still becomes superconducting suggesting that nesting is not necessary for 
superconductivity to develop. Similarly, ARPES experiments on 
superconducting $A$Fe$_{1.6+x}$Se$_2$ \cite{yzhang,tqian,Dxmou} revealed 
the absence of the hole-like FS's necessary 
for the $\Gamma(0,0)\leftrightarrow M(1,0)$ 
$s^{\pm}$-wave superconductivity.  
Also note that related materials such as LaFePO with a well-nested 
FS also do not order magnetically. Why weak coupling arguments would work in some 
cases and not others? Finally, scanning tunneling microscopy (STM) 
experiments~\cite{tmchuang} on Ca(Fe$_{1-x}$Co$_x$)$_2$As$_2$
shows an exotic ``nematic'' electronic structure, 
similar to those found for strongly coupled copper-oxides.

{\it Additional theoretical results}. In 
fluctuation-exchange approximation studies it was concluded 
that the nesting results are not robust against the addition of self-energy 
corrections \cite{arita09}. Other calculations have suggested 
that magnetic order in pnictides is neither fully localized nor fully itinerant: 
the $J_H$ coupling forms the local moments, while the particular ground state 
is selected by itinerant one-electron interactions~\cite{Johannes09}. Moreover, 
studies of a spin-fermion model for the pnictides~\cite{lv10,wgyin10,shliang11} 
revealed the crucial role played by the Hund's rule coupling and suggested that 
the Fe superconductors are closer kin to manganites, where similar 
spin-fermion models were extensively studied~\cite{dagotto-cmr}, 
than to copper-oxides with regards to their diverse magnetism and incoherent 
normal-state electron transport. 

\begin{figure}[t]
\includegraphics[scale=.48]{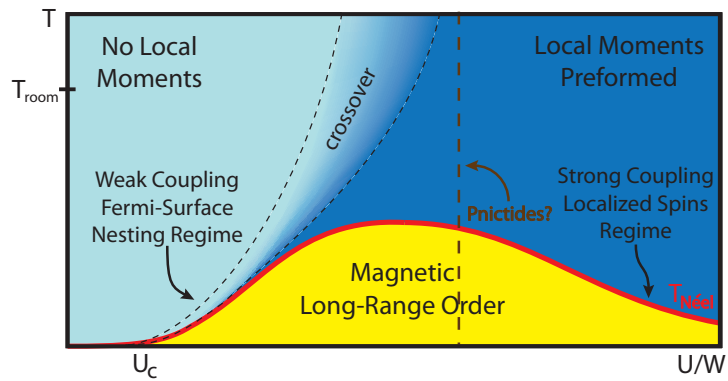}
\caption{
Sketch of the expected phase diagram of the Hubbard model varying temperature 
and $U/W$ in the undoped limit. Highlighted are the regimes of weak coupling 
where nesting dominates and strong coupling where localized spins approaches 
are suitable. At temperatures above $T_N$, there are regions with and without preformed 
local moments. The dashed line tentatively locates the pnictides and chalcogenides 
in the ``middle'', with a physics involving itinerant electrons coexisting with localized 
moments. Note that this phase diagram is guided by results 
known for the one-orbital case, while the true multiorbital Hubbard model phase 
diagram may display an even richer structure. In particular, a second critical $U/W$ 
at low temperatures separating the metallic AF state from the insulating AF state
is not shown for simplicity.}
\end{figure}

\section{Conclusions}

Recent studies of Fe-based superconductors 
are unveiling a perspective of these exciting materials
that is far richer than previously anticipated. While in the early days weak coupling
approaches seemed sufficient to understand
these compounds, several recent efforts, reviewed in part here, suggest 
that understanding the physics of these materials may require more 
refined concepts, better many-body theoretical calculations, and additional 
sophisticated experiments for a more in-depth rationalization 
of their properties. In fact, evidence 
is building that pnictides and chalcogenides inhabit the mostly unexplored 
``intermediate'' region of Hubbard $U/W$ couplings, which is neither very weak 
coupling, where FS nesting concepts apply, nor strong coupling, where localized 
spins provide a good starting point as it occurs in the undoped copper oxides. 
The situation is qualitatively summarized 
in Fig. 4 where a crude sketch of a plausible phase 
diagram for a generic undoped Hubbard model is provided varying temperature $T$ and 
$U/W$ at, e.g., a fixed $J_H/U$ such as 1/4. In weak coupling, first a critical 
value of $U$ must be crossed before magnetic order develops at low temperatures. 
In this region, nesting works properly. As $U$ increases, $T_N$ first 
increases, reaches 
a broad maximum, and then eventually in the regime of localized spins $T_N$ starts to 
decrease since it becomes regulated by the Heisenberg superexchange that scales 
as $1/U$. Above $T_N$ a ``crossover'' temperature that roughly grows like $U$ is 
shown separating regions with and without ``preformed'' local moments. Since the 
pnictides have local moments at room temperature, 
then a tentative location 
for these materials is provided by the dashed line. 
However, whether this line coincides with the 
maximum $T_N$ or is shifted to the left or the right is too early to say, but it 
cannot be too far from optimal otherwise local moments would be absent, if far left, 
or an insulator should be found at low temperatures, if far right. 
Theoretical mean-field
estimates reviewed here using the multiorbital Hubbard model find that $U/W$$\sim$$0.3-0.5$
could work for pnictides. However, for chalcogenides and alkaline iron selenides, 
and also after including quantum fluctuations, the ratio $U/W$ may increase further, 
and it may reach the $U/W\sim1$ threshold widely consider to mark the starting point for a strong coupling
description. Note also that the 
sketch in Fig.~4 is based on our knowledge on the one-orbital 
Hubbard model and a proper multiorbital analysis will lead to an even richer phase diagram. 
In fact, a critical $U$ for the transition between the magnetic metallic state 
and the magnetic insulating state at low temperatures 
should also be present, but it is 
not shown in the sketch for simplicity: this transition should occur at a 
$U$ larger than the pnictides dashed line since these materials are metallic at low temperatures.

In summary, the Fe-based superconductors continue surprising us with their
exotic properties that do not fit into the simple limits of weak or strong coupling $U$.
Additional experimental and theoretical efforts are
needed to unveil the secrets of this intriguing family of materials.


\begin{thebibliography}{}
\bibitem{bednorz} Bednorz, J. G. \& M$\rm \ddot{u}$ller, K. A., 
Possible high-$T_c$ superconductivity in the Ba-La-Cu-O system.  Z. Phys. B {\bf 64}, 189-193 (1986).
\bibitem{vaknin} Vaknin, D. {\it et al.}, Antiferromagnetism in La$_2$CuO$_{4-y}$. Phys. Rev. Lett. {\bf 58}, 2802-2805 (1987).
\bibitem{tranquada88} Tranquada, J. M. {\it et al.}, Neutron-Diffraction Determination of Antiferromagnetic Structure of Cu Ions in YBa$_2$Cu$_3$O$_{6+x}$ with $x = 0.0$ and 0.15. Phys. Rev. Lett. {\bf 60}, 156-159 (1988).
\bibitem{scalapino} Scalapino, D. J., The case for $d_{x^2-y^2}$ pairing in the cuprate superconductors.
Physics Reports {\bf 250}, 329-365 (1995).
\bibitem{dagotto} Dagotto, E., Correlated electrons in high-temperature superconductors. 
Rev. Mod. Phys. {\bf 66}, 763 (1994).
\bibitem{palee} Lee, P. A., Nagaosa, N., and Wen, X.-G., Doping a Mott insulator: Physics of high-temperature superconductivity. 
Rev. Mod. Phys. {\bf 78}, 17-85 (2006).
\bibitem{fujita} Fujita, M. {\it et al.}, Progress in Neutron Scattering Studies of
Spin Excitations in High-$T_c$ Cuprates. J. Phys. Soc. Jpn. {\bf 81}, 011007 (2012).
\bibitem{johnston} Johnston, D. C., The Puzzle of High Temperature Superconductivity in Layered Iron Pnictides and Chalcogenides. 
Advances in Physics {\bf 59}, 803-1061 (2010).
\bibitem{stewart} Stewart, G. R., Superconductivity in iron compounds. Rev. Mod. Phys. {\bf 83}, 1589-1652 (2011).
\bibitem{paglione10} Paglione, J. \& Greene, R. L., High-temperature superconductivity in
iron-based materials. Nature Physics {\bf 6}, 645-658 (2010).
\bibitem{kamihara} Kamihara, Y., Watanabe, T., Hirano, M. and Hosono, H. Iron-based layered
superconductor La[O$_{1-x}$F$_x$]FeAs ($x=0.05$-0.12) with $T_c=26$ K, J. Am. Chem. Soc. {\bf 130}, 3296-3297 (2008).
\bibitem{rotter} Rotter, M., Tegel, M., \& Johrendt, D., 
Superconductivity at 38 K in the Iron Arsenide (Ba$_{1-x}$K$_x$)Fe$_2$As$_2$. 
Phys. Rev. Lett. {\bf 101}, 107006 (2008).
\bibitem{cwchu} Chu, C. W., {\it et al.}, The synthesis and characterization of
LiFeAs and NaFeAs. Physica C {\bf 469}, 326-331 (2009).
\bibitem{mawkuen2} Hsu, F.-C. {\it et al.}, 
Superconductivity in the PbO-type structure $\alpha$-FeS. 
Proc. Natl. Acad. Sci. U.S.A {\bf 105}, 14262 (2008).
\bibitem{mazin2011n} Mazin, I. I., Superconductivity gets an iron boost. Nature {\bf 464}, 183-186 (2010).
\bibitem{hirschfeld} Hirschfeld, P. J., Korshunov, M. M., and Mazin, I. I., Gap symmetry and structure of Fe-based superconductors, 
Rep. Prog. Phys. {\bf 74}, 124508 (2011).
\bibitem{dong} Dong, J. {\it et al.}, Competing orders and spin-density-wave instability in LaO$_{1-x}$F$_x$FeAs. Euro. Phys. Lett.
{\bf 83}, 27006 (2008).
\bibitem{fawcett} Fawcett, E.,  Spin-density-wave antiferromagnetism in chromium, 
Rev. Mod. Phys. {\bf 60}, 209-283 (1998).
\bibitem{cruz} de la Cruz, C. {\it et al.}, Magnetic order close to superconductivity in the iron-based layered LaO$_{1-x}$F$_x$FeAs systems,
 Nature \textbf{453}, 899-902 (2008).
\bibitem{qhuang} Huang, Q. {\it et al.}, Neutron-Diffraction Measurements of Magnetic Order and a Structural Transition in the Parent 
BaFe$_2$As$_2$ Compound of FeAs-Based High-Temperature Superconductors. Phys. Rev. Lett. {\bf 101}, 257003 (2008).
\bibitem{slli09} Li, S., {\it et al.}, Structural and magnetic phase transitions in Na$_{1-\delta}$FeAs.
Phys. Rev. B {\bf 80}, 020504(R) (2009).
\bibitem{mazin08} Mazin, I. I., Johannes, M. D., Boeri, L., Koepernik, K., \& Singh, D. J., Problems 
with reconciling density functional theory calculations with experiment in ferropnictides, Phys. Rev. B {\bf 78}, 085104 (2008).
\bibitem{kuroki08} Kuroki, K. {\it et al.}, Unconventional pairing originating from the disconnected Fermi surfaces of superconducting LaFeAsO$_{1-x}$F$_x$. Phys. Rev. Lett. {\bf 101}, 087004 (2008).
\bibitem{chubukov} Chubukov, A. V., Pairing mechanism in
Fe-based superconductors. Annu. Rev. Condens. Matter Phys. {\bf 3}, 13 (2012). 
\bibitem{fwang09} Wang, F., \& Lee, D.-H., The electron-pairing mechanism of
iron-based superconductors. Science {\bf 332}, 200-204 (2011).
\bibitem{eschrig} Eschrig, M., The effect of collective spin-1 excitations on electronic spectra in high-$T_c$ superconductors. Advances in Physics {\bf 55}, 47-183 (2006).
\bibitem{maier} Maier, T. A., and Scalapino, D. J., Theory of neutron scattering as a probe of the superconducting gap in the iron pnictides.
Phys. Rev. B {\bf 78}, 020514(R) (2008).
\bibitem{korshunov} Korshunov, M. M. and Eremin, I., Theory of magnetic excitations in 
iron-based layered superconductors. Phys. Rev. B {\bf 78}, 140509(R) (2008).
\bibitem{christianson} Christianson, A. D. {\it et al.}, 
Resonant Spin Excitation in the High Temperature Superconductor Ba$_{0.6}$K$_{0.4}$Fe$_2$As$_2$.
Nature {\bf 456}, 930-932 (2008).
\bibitem{chenglinzhang} Zhang, C. L. {\it et al.}, Neutron scattering studies of spin
excitations in hole-doped Ba$_{0.67}$K$_{0.33}$Fe$_2$As$_2$ superconductor. Scientific Reports {\bf 1}, 115 (2011).
\bibitem{castellan} Castellan, J.-P. {\it et al.}, Effect of Fermi surface nesting on resonant spin excitations in Ba$_{1-x}$K$_x$Fe$_2$As$_2$.
Phys. Rev. Lett. {\bf 107}, 177003 (2011).
\bibitem{lumsden} Lumsden, M. D. {\it et al.,} Two-dimensional resonant magnetic excitation in BaFe$_{1.84}$Co$_{0.16}$As$_2$. Phys. Rev. Lett. {\bf 102}, 107005 (2009).
\bibitem{schi09} Chi, S. {\it et al.,} Inelastic neutron-scattering measurements of a three-dimensional spin resonance in the FeAs-based BaFe$_{1.9}$Ni$_{0.1}$As$_2$ superconductor. 
Phys. Rev. Lett. {\bf 102}, 107006 (2009).
\bibitem{dsinosov09}	Inosov, D. S. {\it et al.,} Normal-state spin dynamics and temperature-dependent spin resonance energy in an optimally doped iron arsenide superconductor. Nat. Phys. {\bf 6}, 178 (2010).
\bibitem{jtpark} Park, J. T. {\it et al.,} Symmetry of spin excitation spectra in tetragonal paramagnetic and superconducting phases of 122-ferropnictides. Phys. Rev. B {\bf 82}, 134503 (2010).
\bibitem{clester} Lester, C. {\it et al.}, Dispersive spin fluctuations in the nearly optimally doped superconductor Ba(Fe$_{1-x}$Co$_x$)$_2$As$_2$ ($x = 0.065$). Phys. Rev. B {\bf 81}, 064505 (2010).
\bibitem{hfli} Li, H. F. {\it et al.}, Anisotropic and quasipropagating spin excitations in superconducting Ba(Fe$_{0.926}$Co$_{0.074}$)$_2$As$_2$. Phys. Rev. B {\bf 82}, 140503(R) (2010).
\bibitem{hqluo12} Luo, H. Q. {\it et al.}, Electron doping evolution of the anisotropic spin excitations in BaFe$_{2-x}$Ni$_x$As$_2$. Phys. Rev. B {\bf 86}, 024508 (2012).
\bibitem{hamook} Mook, H. A. {\it et al.}, Unusual relationship between magnetism and superconductivity in FeTe$_{0.5}$Se$_{0.5}$. Phys. Rev. Lett. {\bf 104}, 187002 (2010).
\bibitem{qiu09} Qiu, Y., {\it et al.}, Spin gap and resonance at the nesting wave vector 
in superconducting FeSe$_{0.4}$Te$_{0.6}$. Phys. Rev. Lett. {\bf 103}, 067008 (2009). 
\bibitem{lumsden2} Lumsden, M. D. {\it et al.}, 
Evolution of spin excitations into the superconducting state in FeTe$_{1-x}$Se$_x$.
Nat. Phys. {\bf 6}, 182-186 (2010).
\bibitem{prichard} Richard, P., Sato, T., Nakayama, K., Takahashi, T.,
Ding, H., Fe-based superconductors: an angle-resolved photoemission spectroscopy perspective. Report on Progress in Physics {\bf 74}, 124512 (2011). 
\bibitem{si2008} Si. Q., and Abrahams, E., Strong correlations and magnetic frustration in the high $T_c$ iron pnictides. 
Phys. Rev. Lett. {\bf 101}, 076401 (2008).
\bibitem{qmsi09} Si, Q., Abrahams, E., Dai, J. H., Zhu, J.-X., Correlation effects in the iron pnictides, New J. Phys. {\bf 11}, 045001 (2009). 
\bibitem{cfang} Fang, C., Yao, H., Tsai, W. F., Hu, J. P. \& Kivelson, S. A. Theory of electron
nematic order in LaOFeAs. Phys. Rev. B 77, 224509 (2008).
\bibitem{cxu} Xu, C. K., M$\rm \ddot{u}$ller, \& Sachdev, S., 
Ising and spin orders in the iron-based superconductors. 
Phys. Rev. B {\bf 78}, 020501(R) (2008).
\bibitem{seo2008}
  Seo, K.,  Bernevig B. A.  \& Hu, J. P. Pairing symmetry in a two-orbital exchange coupling model of oxypnictides. \textit{Phys. Rev. Lett.} {\bf
101},  206404  (2008).
\bibitem{Fang2011}
Fang, C. {\it et al.} Robustness of $s$-wave pairing in electron overdoped A$_{1-y}$Fe$_{2-x}$Se$_2$.  \textit{Phy. Rev. X} \textbf{ 1}, 011009 (2011)
\bibitem{nicholson11}
Nicholson, A., {\it et al.}, 
Competing Pairing Symmetries in a Generalized Two-Orbital Model for the Pnictide Superconductors,
Phys. Rev. Lett. {\bf 106}, 217002 (2011).
\bibitem{jgguo} Guo, J. G. {\it et al}., Superconductivity in the iron selenide K$_x$Fe$_2$Se$_2$ ($0\leq x\leq 1.0$). Phys. Rev. B {\bf 82}, 180520(R) (2010).
\bibitem{mhfang} Fang, M. H. {\it et al.}, Fe-based high temperature superconductivity with $T_c=31$ K 
bordering an insulating antiferromagnet in (Tl,K)Fe$_x$Se$_2$ Crystals. EuroPhys. Lett. {\bf 94}, 27009 (2011).
\bibitem{Wang_122Se}
Wang, X.-P. {\it et al.} Strong nodeless pairing on separate electron Fermi surface sheets in (Tl,K)Fe$_{1.78}$Se$_2$ probed by ARPES. \textit{Europhys. Lett.} \textbf{93}, 57001 (2011).
\bibitem{Zhang_122Se}
Zhang, Y. {\it et al}.  Heavily electron-doped electronic structure and isotropic superconducting gap in A$_x$Fe$_2$Se$_2$ (A=K,Cs).   \textit{Nature Mater.} \textbf{10}, 273-277 (2011).
\bibitem{Mou_122Se}
Mou, D.  {\it et al.} Distinct Fermi surface topology and nodeless superconducting gap in a (Tl$_{0.58}$Rb$_{0.42}$)Fe$_{1.72}$Se$_2$ superconductor. \textit{Phys. Rev. Lett.} \textbf{106}, 107001 (2011).
\bibitem{wbao1} Bao, W. {\it et al.}
A Novel Large Moment Antiferromagnetic Order in K$_{0.8}$Fe$_{1.6}$Se$_2$ Superconductor. 
Chinese Phys. Lett. {\bf 28}, 086104 (2011).
\bibitem{fye} Ye, F., {\it et al.}, Common crystalline and magnetic structure of superconducting 
$A_2$Fe$_4$Se$_5$ ($A=$ K,Rb,Cs,Tl) single crystals measured using neutron diffraction.
Phys. Rev. Lett. {\bf 107}, 137003 (2011).
\bibitem{qazilbash09} Qazilbash, M. M., Hamlin, J. J., 
Baumbach, R. E., Zhang, L. J., Singh, D. J., Maple, M. B., Basov, D. N.,  Electronic correlations in the iron pnictides, Nature Physics {\bf 5}, 647 (2009). 
\bibitem{HG} Haule, K., Shim, J. H., \& Kotliar, G.,  
Correlated electronic structure of LaO$_{1-x}$F$_x$FeAs.
Phys. Rev. Lett. {\bf 100}, 226402 (2008).
\bibitem{coldea} Coldea, R. {\it et al}. Spin waves and electronic interactions in La$_2$CuO$_4$.
Phys. Rev. Lett. {\bf 86}, 5377-5380 (2001).
\bibitem{headings} Headings, N. S., Hayden, S. M., Coldea, R., \& Perring, T. G.,
Anomalous high-energy spin excitations in the high-$T_c$ superconductor-parent antiferromagnet 
La$_2$CuO$_4$. Phys. Rev. Lett. {\bf 105}, 247001 (2010).
\bibitem{mhfang08} Fang, M. H. {\it et al.}, Superconductivity close to magnetic instability in Fe(Se$_{1-x}$Te$_x$)$_{0.82}$.
Phys. Rev. B {\bf 78}, 224503 (2008).

\bibitem{subedi} Subedi, A., Zhang, L. J., Dingh, D. J., \& Du, M. H., 
Density functional study of FeS, FeSe, and FeTe: electronic structure, magnetism, phonons, and superconductivity. Phys. Rev. B {\bf 78}, 134514 (2008).
\bibitem{webao08} Bao, W., {\it et al.}, Tunable $(\delta\pi,\delta\pi)$-type antiferromagnetic order in $\alpha$-Fe(Te,Se) superconductors.
Phys. Rev. Lett. {\bf 102}, 247001 (2009). 
\bibitem{slli08} Li, S. L., {\it et al.}, First-order magnetic and structural phase transitions in Fe$_{1+y}$Se$_x$Te$_{1-x}$. Phys. Rev. B {\bf 79}, 054503 (2009).

\bibitem{diallo09} Diallo, S. O. {\it et al.}, Itinerant magnetic excitations in antiferromagnetic CaFe$_2$As$_2$.
Phys. Rev. Lett. {\bf 102}, 187206 (2009).
\bibitem{jzhao} Zhao, J. {\it et al.}, Spin waves and magnetic exchange interactions in CaFe$_2$As$_2$. 
Nat. Phys. {\bf 5}, 555-560 (2009).
\bibitem{raewings} Ewings, R. A. {\it et al.}, 
Itinerant spin excitations in SrFe$_2$As$_2$ measured by inelastic neutron scattering. Phys. Rev. B {\bf 83},
214519 (2011).
\bibitem{lharriger} Harriger, L. W. {\it et al.}, Nematic spin fluid in the tetragonal phase of BaFe$_2$As$_2$. Phys. Rev. B {\bf 84}, 054544 (2011).
\bibitem{rodriguez} Rodriguez, E. E. {\it et al.}, Magnetic-crystallographic 
phase diagram of the superconducting parent compound Fe$_{1+x}$Te. Phys. Rev. B {\bf 84},
064403 (2011).
\bibitem{lipscombe} Lipscombe, O. J. {\it et al.}, Spin waves in the 
$(\pi,0)$ magnetically ordered iron chalcogenide Fe$_{1.05}$Te. Phys. Rev. Lett. {\bf 106}, 057004 (2011).
\bibitem{zaliznyak} Zaliznyak, I. A., {\it et al.}, 
Unconventional temperature enhanced magnetism in iron telluride. 
Phys. Rev. Lett. {\bf 107}, 216403 (2011).
\bibitem{mywang11} Wang, M. Y., {\it et al.}, Spin waves and magnetic exchange interactions in insulating Rb$_{0.89}$Fe$_{1.58}$Se$_2$. Nature Communications {\bf 2}, 580 (2011).
\bibitem{nni08} Ni, N., {\it et al.}, Effects of Co substitution on thermodynamic and transport properties and anisotropic $H_{c2}$ in Ba(Fe$_{1-x}$Co$_x$)$_2$As$_2$ single crystals. Phys. Rev. B {\bf 78}, 214515 (2008).
\bibitem{jhchu09} Chu, J.-H., {\it et al.}, Determination of the phase diagram of the electron-doped superconductor Ba(Fe$_{1-x}$Co$_x$)$_2$As$_2$.
Phys. Rev. B {\bf 79}, 014506 (2009).
\bibitem{lester09} Lester, C., {\it et al.}, Neutron scattering study of the interplay between structure and magnetism in Ba(Fe$_{1-x}$Co$_x$)$_2$As$_2$.
Phys. Rev. B {\bf 79}, 144523 (2009). 
\bibitem{pratt09} Pratt, D. K., {\it et al.}, Coexistence of competing antiferromagnetic and superconducting phases in the underdoped Ba(Fe$_{0.953}$Co$_{0.047}$)$_2$As$_2$ compound using X-ray and neutron scattering techniques. Phys. Rev. Lett. {\bf 103}, 087001 (2009).
\bibitem{adchristianson09} Christianson, A. D., {\it et al.}, Static and dynamic magnetism in underdoped superconductor BaFe$_{1.92}$Co$_{0.08}$As$_2$. 
Phys. Rev. Lett. {\bf 103}, 087002 (2009).
\bibitem{mywang} Wang, M. Y. {\it et al.}, Electron-doping evolution of the low-energy spin excitations in the iron arsenide superconductor BaFe$_{2-x}$Ni$_x$As$_2$. Phys. Rev. B {\bf 81}, 174524 (2010).
\bibitem{mywang11b} Wang, M. Y. {\it et al.}, Magnetic field effect on static antiferromagnetic order and spin excitations in the underdoped iron arsenide superconductor BaFe$_{1.92}$Ni$_{0.08}$As$_2$. Phys. Rev. B {\bf 83}, 094516 (2011).
\bibitem{dkpratt11} Pratt, D. K., {\it et al.}, Incommensurate spin-density wave order in electron-doped BaFe$_2$As$_2$ superconductors.
Phys. Rev. Lett. {\bf 106}, 257001 (2011).
\bibitem{hluo12} Luo, H. Q., {\it et al.}, Coexistence and competition of the short-Range incommensurate antiferromagnetic order with the superconducting state of BaFe$_{2-x}$Ni$_x$As$_2$. Phys. Rev. Lett. 
{\bf 108}, 247002 (2012).   
\bibitem{snandi10} Nandi, S., {\it et al.}, Anomalous suppression of the orthorhombic lattice distortion in 
superconducting Ba(Fe$_{1-x}$Co$_x$)$_2$As$_2$ Single Crystals. Phys. Rev. Lett. {\bf 104}, 057006 (2010).
\bibitem{hchen09} Chen, H., {\it et al.}, Coexistence of the spin-density wave and superconductivity
in Ba$_{1-x}$K$_x$Fe$_2$As$_2$. EPL {\bf 85}, 17006 (2009).
\bibitem{jtpark09} Park, J. T., {\it et al.}, Electronic phase separation in the slightly underdoped
iron pnictide superconductor Ba$_{1-x}$K$_x$Fe$_2$As$_2$. Phys. Rev. Lett. {\bf 102}, 117006 (2009).
\bibitem{savci} Avci, S. {\it et al.}, Magnetoelastic coupling in the phase diagram of Ba$_{1-x}$K$_x$Fe$_2$As$_2$ as seen via neutron diffraction.
Phys. Rev. B {\bf 83}, 172503 (2011).
\bibitem{ewiesenmayer11} Wiesenmayer, E., {\it et al.}, Microscopic coexistence of superconductivity and magnetism in Ba$_{1-x}$K$_x$Fe$_2$As$_2$.
Phys. Rev. Lett. {\bf 107}, 237001 (2011).
\bibitem{graser10} Graser, S. {\it et al.}, Spin fluctuations and superconductivity in a three-dimensional tight-binding model for BaFe$_2$As$_2$.
Phys. Rev. B {\bf 81}, 214503 (2010).
\bibitem{jhzhang10} Zhang, J. H., Sknepnek, R., \& Schmalian, J., Spectral analysis for the iron-based superconductors: Anisotropic spin fluctuations and fully
gapped $s^{\pm}$-wave superconductivity. Phys. Rev. B {\bf 82}, 134527 (2010).
\bibitem{msliu12} Liu, M. S. {\it et al.}, Nature of magnetic excitations in 
superconducting BaFe$_{1.9}$Ni$_{0.1}$As$_2$. Nat. Phys. {\bf 8}, 376-381 (2012).
\bibitem{chlee11} Lee, C. H., {\it et al.}, Incommensurate spin fluctuations in hole-overdoped superconductor KFe$_2$As$_2$. Phys. Rev. Lett. {\bf 106}, 067003 (2011).
\bibitem{hpark12} Park, H., Haule, K., \& Kotliar, G.,
Magnetic excitation spectra in BaFe$_2$As$_2$: a two-particle approach within a combination of the density functional theory and the dynamical mean-field theory method. Phys. Rev. Lett. {\bf 107}, 137007 (2011).
\bibitem{tterashima10} Terashima, T., {\it et al.}, Fermi surface and mass enhancement in KFe$_2$As$_2$ from de Haas-van Alphen effect measurements, 
J. Phys. Soc. Jpn. {\bf 79}, 053702 (2010).
\bibitem{pmcrourke} Rourke, P. M. C. {\it et al.}, A detailed de Haas����Cvan Alphen effect study of the
overdoped cuprate Tl$_2$Ba$_2$CuO$_{6+\delta}$. New Journal of Physics {\bf 12}, 105009 (2010). 
\bibitem{nakamura} Nakamura, K., Arita, R., \& Imada, M., J. Phys. Soc. Jpn. {\bf 77}, 093711 (2008).
\bibitem{dsinosov11} Inosov, D. S., {\it et al.,}  
Crossover from weak to strong pairing in unconventional superconductors. Phys. Rev. B {\bf 83}, 214520 (2011).
\bibitem{zpyin10} Yin, Z. P., Haule, K., \& Kotliar, G., Kinetic frustration and the nature of the magnetic and paramagnetic states in iron pnictides and iron chalcogenides.
 Nature Materials {\bf 10}, 932-935 (2011).
\bibitem{che10} He, C., {\it et al.}, Electronic-structure-driven magnetic and structure transitions in superconducting NaFeAs single crystals measured by Angle-Resolved Photoemission Spectroscopy, Phys. Rev. Lett. {\bf 105}, 117002 (2010).
\bibitem{qluo10} Luo, Q., {\it et al.},  Neutron and ARPES constraints on the couplings of the multiorbital Hubbard model for the iron pnictides, 
Phys. Rev. B {\bf 82}, 104508 (2010).
\bibitem{mDaghofer}
Daghofer, M., Nicholson, A., Moreo, A., \& Dagotto, E., 
Three orbital model for the iron-based superconductors, Phys. Rev. B {\bf 81}, 014511 (2010).
\bibitem{kkubo11} Kubo, K. \& Thalmeier, P., Correlation effects on antiferromagnetism in Fe pnictides.
J. Phys. Soc. Jpn. {\bf 80}, SA121 (2011).
\bibitem{arita-imada} Miyake, T., Nakamura, K., Arita, R., \& Imada, M., Comparison of Ab initio Low-Energy Models
for LaFePO, BaFe2As2, LiFeAs, FeSe, and FeTe: Electron Correlation and Covalency, J. Phys. Soc. Jpn. {\bf 79}, 044705 (2010). 
\bibitem{Johannes09} Johannes, M. D. \& Mazin, I. I., Microscopic origin of magnetism and magnetic interactions in ferropnictides, 
Phys. Rev. B {\bf 79}, 220510(R) (2009).
\bibitem{hund-metal} Haule, K. \& Kotliar, G., 
Coherence-incoherence crossover in the normal state of iron oxypnictides and importance of Hund's rule coupling.
New J. of Phys. {\bf 11}, 025021 (2009).
\bibitem{lv10} Lv, W. L., Kr$\rm \ddot{u}$ger, F., \& Phillips, P., Orbital ordering and unfrustrated $(\pi,0)$ magnetism from degenerate double exchange 
in the iron pnictides. Phys. Rev. B {\bf 82}, 045125 (2010).
\bibitem{wgyin10} Yin, W.-G., Lee, C. C., \& Ku, W., Unified picture for magnetic correlations in iron-based superconductors, Phys. Rev. Lett. {\bf 105}, 107004 (2010).
\bibitem{shimojima} Shimojima, T., {\it et al.}, Orbital-dependent modifications of electronic structure across the magnetostructural transition in 
BaFe$_2$As$_2$. Phys. Rev. Lett. {\bf 104}, 057002 (2010). 
\bibitem{daghofer10} Daghofer, M., {\it et al.,} Orbital-weight redistribution triggered by spin order in the pnictides, Phys. Rev. B {\bf 81}, 180514(R) (2010).
\bibitem{daghofer12} Daghofer, M., Nicholson, A., and Moreo, A., Spectral density in a nematic state of iron pnictides, Phys. Rev. B {\bf 85}, 184515 (2012).
\bibitem{irfisher11} Fisher, I. R., Degiorgi, L., \& Shen, Z. X., In-plane electronic anisotropy of
underdoped `122' Fe-arsenide superconductors revealed by measurements of detwinned single crystals. Rep. Prog. Phys. {\bf 74}, 124506 (2011).
\bibitem{matanatar} Tanatar, M. A., {\it et al.}, Uniaxial-strain mechanical detwinning of CaFe$_2$As$_2$ and BaFe$_2$As$_2$ crystals: Optical and transport study. Phys. Rev. B {\bf 81}, 814508 (2010).
\bibitem{xtzhang} Zhang, X. T., \& Dagotto, E., Anisotropy of the optical conductivity of a pnictide superconductor from the undoped three-orbital Hubbard model, Phys. Rev. B {\bf 84}, 132505 (2011). 
\bibitem{efradkin} Fradkin, E., Kivelson, S. A., Lawler, M. J., Eisenstein, J. P., \& Mackenzie, A. P., Nematic Fermi fluids in condensed matter physics.
 Annu. Rev. Condens. Matter Phys. {\bf 1}, 153-178 (2010). 
\bibitem{myi} Yi, M., {\it et al.}, Symmetry breaking orbital anisotropy on detwinned Ba(Fe$_{1-x}$Co$_x$)$_2$As$_2$ above the spin density wave transition, PNAS {\bf 108}, 6878 (2011). 
\bibitem{Nakajima} Nakajima, M., {\it et al.}, Unprecedented anisotropic metallic state in undoped iron arsenide BaFe$_2$As$_2$ revealed by optical spectroscopy, PNAS {\bf 108}, 12238 (2011). 
\bibitem{dhital} Dhital, C. {\it et al.}, Effect of uniaxial strain on the structural and magnetic phase transitions in BaFe$_2$As$_2$. Phys. Rev. Lett. {\bf 108}, 087001 (2012). 
\bibitem{kasahara} Kasahara, S. {\it et al.}, Electronic nematicity above the structural and 
superconducting transition in BaFe$_2$(As$_{1-x}$P$_x$)$_2$. Nature {\bf 486}, 382-385 (2012).
\bibitem{rmfernandes} Fernandes, R. M., Chubukov, A. V., Knolle, J., Eremin, I., Schmalian, J., 
Preemptive nematic order, pseudogap, and orbital order in the iron pnictides. Phys. Rev. B {\bf 85}, 024534 (2012).
\bibitem{hgretarsson} Gretarsson H., {\it et al.}, Revealing the dual nature of magnetism in iron pnictides and iron chalcogenides using x-ray emission spectroscopy.
Phys. Rev. B {\bf 84}, 100509 (2011).
\bibitem{bondino} Bondino, F., {\it et al.,} Evidence for strong itinerant spin fluctuations in the normal state of CeFeAsO$_{0.89}$F$_{0.11}$ iron-oxypnictide superconductors. Phys. Rev. Lett. {\bf 101}, 267001 (2008).
\bibitem{hansmann} Hansmann, P., {\it et al.,} Dichotomy between large local and small ordered magnetic moments in iron-based superconductors, 
Phys. Rev. Lett. {\bf 104}, 197002 (2010).
\bibitem{shimojima11} Shimojima, T., {\it et al.,} Orbital-independent superconducting gaps in iron pnictides, Science {\bf 332}, 564 (2011).
\bibitem{amoreo} Moreo, A., {\it et al.},
Properties of a two-orbital model for oxypnictide superconductors: Magnetic order, $B_{2g}$ spin-singlet pairing channel, and its nodal structure, Phys. Rev. B {\bf 79}, 134502 (2009). 
\bibitem{nicholson} Nicholson, 
A., {\it et al.},
Role of degeneracy, hybridization, and nesting in the properties of multi-orbital systems, Phys. Rev. B {\bf 84}, 094519 (2011).
\bibitem{bjarnold} Arnold, B. J., {\it et al.}, Nesting of electron and hole Fermi surfaces in nonsuperconducting BaFe$_2$P$_2$, 
Phys. Rev. B {\bf 83}, 220504 (2011).
\bibitem{Borisenko} Borisenko, S. V., {\it et al.}, Superconductivity without Nesting in LiFeAs,  Phys. Rev. Lett. {\bf 105}, 067002 (2010).
\bibitem{tmchuang} Chuang, T.-M., {\it et al.}, Nematic electronic structure in the ``Parent'' state of the iron-based superconductor Ca(Fe$_{1-x}$Co$_x$)$_2$As$_2$, 
Science {\bf 327}, 181 (2010).
\bibitem{yzhang} Zhang,Y. {\it et al.},
Heavily electron-doped electronic structure and isotropic superconducting gap in $A_x$Fe$_2$Se$_2$ ($A=$K,Cs), 
Nature Materials {\bf 10}, 273-277 (2011).
\bibitem{tqian} Qian, T. {\it et al.},  
Absence of holelike Fermi surface in superconducting K$_{0.8}$Fe$_{1.7}$Se$_2$ revealed by ARPES,
Phys. Rev. Lett. {\bf 106}, 187001 (2011).
\bibitem{Dxmou} Mou, D. X. {\it et al.}, Distinct Fermi Surface Topology and Nodeless Superconducting Gap in 
(Tl$_{0.58}$Rb$_{0.42}$)Fe$_{1.72}$Se$_2$ Superconductor. Phys. Rev. Lett. {\bf 106}, 107001 (2011).
\bibitem{arita09} Arita, R., \& Ikeda, H., Is Fermi-surface nesting the origin of superconductivity in iron pnictides?: a fluctuation-exchange-approximation study.
J. Phys. Soc. Jpn. {\bf 78}, 113707 (2009).
\bibitem{dagotto-cmr} Dagotto, E., Hotta, T., \& Moreo, A., Colossal magnetoresistant materials: the key role of phase separation, Physics Reports {\bf 344}, 1 (2001).




\end{thebibliography}

\begin{flushleft}
{\bf Acknowledgments} 
We thank Leland W. Harriger for preparing the figures shown in this manuscript. We are also grateful to Thomas A. Maier for calculating the Fermi surfaces of BaFe$_2$As$_2$ shown in Fig. 2d. 
P.D. is supported by the U.S. NSF DMR-1063866, OISE-0968226, and by U.S. DOE,
BES, under Grant 
No. DE-FG02-05ER46202. Work at Institute of Physics is supported by the
Ministry of Science and Technology of China 973 program (2012CB821400). 
E.D. is supported by the U.S. DOE, BES, Materials Sciences and Engineering
Division, and by the U.S. NSF DMR-11-04386.
\end{flushleft}

\begin{flushleft}
{\bf Author contributions} 
\end{flushleft}
P.D. and E.D. wrote the experimental and theoretical portions of the article, respectively. 
J.P.H. revised the article.  All authors discussed the outline of the article.
\begin{flushleft}
{\bf Additional information} 
The authors declare no competing financial interests. 

{\bf Reprints and permissions}
information is available online at http://ngp.nature.com/reprintsandpermissions/. 

Correspondence and
requests for materials should be addressed to P.D. (e-mail: pdai@utk.edu) or E.D. (e-mail: edagotto@utk.edu).
\end{flushleft}

\end{document}